\documentclass[conference]{IEEEtran}

\usepackage{fullpage}
\usepackage{amssymb}
\usepackage{siunitx}
\usepackage{amsmath}
\usepackage{relsize}
\usepackage{mathtools}
\usepackage{verbatim}
\usepackage{graphicx}

\def\multiset#1#2{\ensuremath{\left(\kern-.3em\left(\genfrac{}{}{0pt}{}{#1}{#2}\right)\kern-.3em\right)}}

\usepackage{afterpage}

\begin{document}

\title{The Effect of Erasure Coding on the Burstiness of Packet Loss}

\author{\IEEEauthorblockN{Bobak McCann}
\IEEEauthorblockA{Johns Hopkins Applied Physics Laboratory\\
Laurel, MD\\
Email: Bobak.McCann@jhuapl.edu}
\and
\IEEEauthorblockN{Kerry Fendick}
\IEEEauthorblockA{Johns Hopkins Applied Physics Laboratory\\
Laurel, MD\\
Email: Kerry.Fendick@jhuapl.edu}}

\maketitle

\begin{abstract}
The perceived quality of real-time media delivered over IP networks depends on both the rate and burstiness of packet loss. In this paper, we develop a new mathematical model for the residual burstiness of loss under erasure coding. We derive the expected number of consecutive losses in a burst as a function of erasure coding parameters and the network loss probability assuming a Bernoulli model for network losses. 
\end{abstract}

\section{Introduction}
Erasure coding is a forward error correction scheme used to decrease the probability of packet loss in a network through the use of redudancy packets. For every $N$ packets containing media or data content (which are grouped together into something called a block), an extra $K$ packets are transmitted with the property that the original $N$ packets can be reconstructed if any $N$ out of the $N+K$ packets are received. In that case, the erasure decoder transmits the original N media or data packets downstream towards the receiving endpoint(s). Otherwise, it transmits the subset of those media packets that itself received from the network. An example of such an $(N + K, K)$ erasure coding scheme is a Reed-Solomon code. Typically, $K<N$, so that such an erasure coding scheme is more bandwidth-efficient than simply transmitting each of the original packets multiple times. Clearly, the more redundancy packets created, the greater the probability that we will be able to reconstruct what the $N$ original packets were. A more complete introduction to erasure coding is presented in [1].\\

The work in this paper was motivated by our study of the benefits of end-to-end erasure coding for improving voice quality; see (unpublished) [2] for details. Unlike data applications, media (voice or video) applications are usually tolerant to some level of packet loss, but the quality of received media depends not only on end-to-end loss rates, but also on the burstiness of losses. In particular, back-to-back losses tend to degrade the perception of media quality more than the same number of  losses that are randomly dispersed, since humans can infer what is missing in small gaps in the audio or video that they hear or see. The ITU-T E-model, specified in G.113 and applied in [2], estimates the conversation quality of voice as a function of parameters that include the expected number of consective losses – one measure of loss burstiness. The impact of block erasure coding on that measure of network performance has not been previously studied.

Our results provide a means of analytically obtaining the expected number of the length of consecutive losses for an $(N+K, K)$ erasure code as a function of the network loss rate. While it is possible to estimate the same quantity statistically from a test bed or simulation, an analytical benchmark, as provided by our model, is necessary to understand whether results obtained statistically are representative of block erasure codes in general, or representative of the particular implementation of erasure coding used in the test bed or simulation. Conversely, results obtained statistically can serve as a check on the logic used in constructing the analytical model. The study in [2] describes how results obtained from a test bed and from the model developed here closely agree.

The two most common models for packet loss across links or networks are the Bernoulli model and the Gilbert-Elliot Model. The former model assumes that each packet is lost with the same probability $p$ and that the probabilities that different packets are lost are independent of one another.  The latter model, which is generally regarded as the better model for packet loss across links, assumes that sequential losses transition between two Bernoulli states with respective probabilities $p_1$ and $p_2$, where the burstiness of packet loss is higher than under the Bernoulli model if $p_1 \neq p_2$. Under that model for link losses, however, the one-state Bernoulli model may still reasonably approximate the process of losses across a network for an individual packet stream since the stream will typically traverse multiple links, and the time interval between its packets on individual links will typically be long relative to the time scale at which bursts of link loss occur. We focus on the case where erasure coding is implemented end-to-end, and therefore assume a one-state Bernoulli model for network packet loss.\\

The paper is organized as follows. In Section II, we derive the probability that a given number of packets are lost in a block after erasure coding. Similar analyses are found in [3] and [4]. In Section III, we determine the expected value of the length of consecutive loss for a single block. In Section IV, we extend our model to describe the expected value of the length of consecutive loss for an infinite sequence of blocks. We also show that a very close approximation can be obtained in a relatively small and computationally feasible number of calculations.\\

\section{Probability of Loss After Erasure Coding}
Assume that we have a block $B$ containing $N$ packets, and a set $A$ of redundancy blocks containing $K$ packets. We will first derive the probability that we have $i$ number of packets in our block that were not able to be reconstructed by the receiver even with erasure coding. Let $p$ be the Bernoulli probability that a given packet is lost on the network (before any erasure coding effects). Then the probability that exactly $j$ packets from $B$ are lost in the network is $\binom{N}{j}p^j (1-p)^{N - j}$, and similarly for the number of packets lost from $A$.\\
 
Using this information, we will derive formulas to describe how many packets are lost after erasure coding.\\

\textbf{Definition: }Let $Q(m)$ be the probability that there are $m$ unrecoverable losses in a block.\\

\newtheorem{theorem}{Theorem}
\begin{theorem}If  packets are lost in the network with Bernoulli probability $p$, then:\\

\begin{footnotesize}
\[ Q(i) = \begin{cases} 
          i=0:  \sum_{i=0}^K \binom{N+K}{i}p^i(1-p)^{N+K-i} \\\\
          0<i\leq K:  \binom{N}{i} p^i (1-p)^{N - i}\sum_{j=0}^{i-1}\binom{K}{K - j} p^{K - j}(1-p)^{j} \\\\
          K < i\leq N: \binom{N}{i} p^i (1-p)^{N - i} 
       \end{cases}
\]\end{footnotesize}\\
\end{theorem}
\begin{IEEEproof}
First consider the case where no losses occur in a block B after erasure coding. In that case, 
$Q(0) = \sum_{i=0}^K \text{Probability($i$ out of $N+K$ packets lost)} \\\\
 \indent \hspace{0.37cm}= \sum_{i=0}^K \binom{N+K}{i}p^i(1-p)^{N+K-i}$\\ 
since we can recover from up to $K$ losses. 
Next, for $0< i \leq K$ packets lost from $B$ in the network, we have\\
$Q(i) = \text{(Probability $i$ packets are lost from $B$ in network)}\\
\indent \hspace{1cm}\cdot \sum_{j=0}^{i-1}\text{(Probability $K-j$ packets are lost from $A$)}$
\\\\$ \indent \hspace{0.32cm} = \binom{N}{i} p^i (1-p)^{N - i}\sum_{j=0}^{i-1}\binom{K}{K - j} p^{K - j}(1-p)^{j}$\\
Finally, for $K < i\leq N$ packets lost from $B$ in the network,\\
$Q(i) = \text{(Probability $i$ packets are lost from $B$ in network)}\\
\indent \hspace{1cm}\cdot \sum_{j=0}^{i-1}\text{(Probability $K-j$ packets are lost from $A$)}$
\\\\$ \indent \hspace{0.3cm} = \binom{N}{i} p^i (1-p)^{N - i}$\\
since no matter how many packets we get from $A$, there will still not be enough packets received to reconstruct any of the $i > K$ packets lost in the network.\\ 
\end{IEEEproof}

\newtheorem{corollary}{Corollary}
\begin{corollary}
The probability that any given packet is lost in the network and is unrecoverable is\\
$$p_L = \frac{\sum\limits_{i=1}^{N} iQ(i)}{N}$$
\end{corollary}

\section{Single Block}
For the remaining results in this paper, we assume without further remarks that packets are lost in the network with Bernoulli probability $p$.
\subsection{Initial Definitions and Notation}
In this section, we find the expected value of the length of consecutive packet loss after erasure coding for a single block of size $N$. Later we will extend this to an infinite sequence of blocks.\\

A block $B$ of size $N$ packets can be represented as a series of blanks and X's, where an X in position $i \ (1\leq i \leq N)$ means that the $i^{th}$ packet was lost in the network and not recoverable. An example is illustrated below:\\

\begin{tiny}
\underline{\hspace{0.21cm}} \underline{\hspace{0.21cm}} \underline{\hspace{0.21cm}} \underline{\hspace{0.21cm}}
\underline{\hspace{0.1cm}}\underline{X}\underline{\hspace{0.1cm}}
\underline{\hspace{0.21cm}} \underline{\hspace{0.21cm}}
\underline{\hspace{0.1cm}}\underline{X}\underline{\hspace{0.1cm}} 
\underline{\hspace{0.1cm}}\underline{X}\underline{\hspace{0.1cm}}
\underline{\hspace{0.21cm}} \underline{\hspace{0.21cm}} \underline{\hspace{0.21cm}} \underline{\hspace{0.21cm}}
\underline{\hspace{0.1cm}}\underline{X}\underline{\hspace{0.1cm}}
\underline{\hspace{0.21cm}} \underline{\hspace{0.21cm}}
\underline{\hspace{0.1cm}}\underline{X}\underline{\hspace{0.1cm}}
\underline{\hspace{0.1cm}}\underline{X}\underline{\hspace{0.1cm}}
\underline{\hspace{0.1cm}}\underline{X}\underline{\hspace{0.1cm}}
\underline{\hspace{0.21cm}} \underline{\hspace{0.21cm}}  
\underline{\hspace{0.1cm}}\underline{X}\underline{\hspace{0.1cm}} 
\underline{\hspace{0.1cm}}\underline{X}\underline{\hspace{0.1cm}}\\
\end{tiny}

\textbf{Definition (informal): } A \textit{pattern} of a block is a sequence of blanks and X-marks.\\

\textbf{Definition: }A \textit{single-block loss row} is a sequence of consecutive X's delimited by either blanks (packet not lost) or by the start or end of the block.\\

\textbf{Definition: }Let $a_1, ..., a_j$ be the lengths of single-block loss rows such that the loss row with length $a_1$ occurs in the block before the loss row of length $a_2$ and so on until the last loss row in the block with length $a_j$. Let $s$ (respectively $e)$ equal 0 if the first (resp. last) packet is lost, and 1 if it is received. Then the sequence $(s, a_1, a_2, ..., a_j, e)$ is called a \textit{loss vector}.

For the example above, $B$ is represented by the loss vector $(0, 1, 2, 1, 3, 2, 1)$.\\

\textbf{Remark: }If we have no single-block loss rows, then our block is an all-receipts block represented by the loss vector $(1,1)$ because it is necessary in this case that $s=e=1$.\\ 

\textbf{Definition: }Let $P(s, a_1, a_2, \dotsc, a_j, e)$ be the probability that the loss vector $(s, a_1, a_2, \dotsc, a_j, e)$ occurs.\\

Notice that distinct loss vectors $ (s_1, a_{1_1}, a_{2_1}, ..., a_{j_1}, e_1), (s_2, a_{1_2}, a_{2_2}, ..., a_{j_2}, e_2)$ are mutually exclusive because the loss vector resulting from a given pattern is unique.\\ 

For the rest of Section III, assume without further remark that the block contains at least one loss.

\subsection{Counting The Number of Patterns Possible for a Given Single-Block loss vector}
A pattern corresponding to the given loss vector $(s, a_1, a_2, ..., a_j, e)$ may be pictured as follows:\\

\begin{small}\fbox{\strut\rule{0.55cm}{0pt}} $X(a_1)$ \fbox{\strut\rule{0.55cm}{0pt}} $X(a_2)$ \fbox{\strut\rule{0.55cm}{0pt}} $\cdots $ \fbox{\strut\rule{0.55cm}{0pt}} $X(a_j)$ \fbox{\strut\rule{0.55cm}{0pt}}\end{small}\\

\noindent where the rectangles are spaces that will each contain a certain number of blanks, and each $X(a_i)$ signifies a loss row of back-to-back $X$'s numbering $a_i$. Two patterns with the same loss vector $(s, a_1, a_2, \dotsc, a_j, e)$ are distinguished by the pattern of blanks between the $a_i$'s of the loss vector and the number of blanks before the first loss row, $a_1$, and after the last loss row $a_j$. To build a unique pattern for a given loss vector, we first need to populate the $j-1$ intermediary spaces (those between the $a_i$'s) with at least one blank each (otherwise we would have a different loss vector). We distribute a blank to each of the intermediary spaces in one way. Also, we need to distribute blanks to the beginning (respectively end) of the block if $s$ (resp. $e$) equals 1.\\

Next, to find the number of unique patterns for a given loss vector $(s, a_1, a_2, \dotsc, a_j, e)$, we distribute the remaining $N - (a_1 + a_2 + \dotsb+ a_j) - (j-1) - s - e$ blanks to the spaces. There are $\multiset{j+1 - (1-s) - (1-e)}{N - (a_1 + a_2 + \dotsb + a_j) - (j-1)  - s - e}$ ways to do this, where the double parantheses denotes the multichoose coefficient. Furthermore, the number of patterns with $a_1 + a_2 + \dotsc a_j$ many losses is $\binom{N}{a_1+a_2+\dotsc +a_j}$.\\\\
Since
$P(s, a_1, a_2, \dotsc, a_j, e) \\\indent\hspace{0.1cm}=\frac{\text{number of patterns in a block of size }N\text{ with the loss vector }(s, a_1, a_2, \dotsc, a_j, e) } {\text{number of patterns in a block of size }N\text{ with } a_1 + a_2 + \dotsb + a_j \text{losses}}  \\\indent\hspace{0.4cm}  \cdot  (  \text{Probability } a_1+a_2+\dotsb + a_j \text{ losses} )$,\\ the next theorem follows:\\
\begin{theorem}The probability of having the loss vector $(s, a_1, a_2, \dotsc, a_j, e)$  is given by the following formula:\\
$P(s, a_1, a_2, \dotsc, a_j, e) \\=  \frac{\multiset{j+1 - (1-s) - (1-e)}{N - (a_1 + a_2 + \dotsb + a_j) - (j-1)  - s - e} \cdot Q(a_1 + a_2 + \dotsb + a_j)}{\binom{N}{a_1+a_2+\dotsb+a_j}}$ \\
\end{theorem}

\subsection{Expected Value of the Length of Consecutive Loss in a Single Block}

Now that we have a formula for the probability of a given loss vector occuring in single block, we need to determine which loss vectors are valid for a block $B$ of size $N$. An invalid loss vector would be a loss vector that requires more than $N$ packets.\\

Since at least one blank is required between each pair of loss rows, the number of loss rows in a block, $j$, must satisfy $1\leq j \leq \Bigl\lfloor{\frac{N+1}{2}}\Bigr\rfloor$. For $j$ in that range, a loss vector $(s, a_1, a_2, \dots , a_j, e)$ is valid if and only if  $s +  a_1 +  a_2 +  \dotsb + a_j +  e + (j-1) \leq N$. Let $C_1$ equal the ratio of losses to loss rows for the block. In other words, $C_1$ is the average length of loss rows for a single-block pattern. Given a valid loss vector $(s, a_1, a_2, \dots , a_j, e)$ with at least one loss, we write the conditional expectation of $C_1$ as \\
\begin{small}$E[C_1 | (s, a_1, a_2, \dots , a_j, e) \text{ and } \sum\limits_{k = 1}^{j}a_k \geq 1] = \frac{ a_1+ a_2+ \dotsb+ a_j}{j}$\end{small}.
We therefore obtain the following theorem:\\

\begin{theorem}The expected value of the length of consecutive losses in a single block is given by\\$E[C_1 | \sum\limits_{k = 1}^{j}a_k \geq 1] = \\\frac{1}{1-Q(0)}\sum_{i\in I}  \frac{a_1+a_2+\dotsb+a_j}{j} P(s, a_1, a_2, \dotsc, a_j, e)$\\
where $I=\big\{(s, a_1, a_2, \dotsc, a_j, e) \big| 1\leq j \leq \Bigl\lfloor{\frac{N+1}{2}}\Bigr\rfloor ,\\ 
\indent\hspace{1.2cm}  s +  a_1 +  a_2 +  \dotsb + a_j +  e + (j-1) \leq N \big\}$.\\
We divide by $1-Q(0)$ because we are conditioning on having at least one loss.
\end{theorem}

\subsection{Estimate Of The Size of the Summation Index for a Single Block}
For practical purposes, we wish to know if we are summing over an easily computable number of terms in Theorem 1.

\textbf{Definition: }Let $p_{n,j}$ equal the number of integer partitions of $n$ into $j$ parts.\\

For a given, valid $j$ \big(i.e., $1\leq j\leq \Bigl\lfloor{\frac{N+1}{2}}\Bigr\rfloor$ \big), the number of loss vectors $(s, a_1, a_2, \dotsc, a_j, e)$ is equal to $\sum_{i=1}^{N - (j-1)}{p_{n, j}}$ because $j \leq a_1+a_2+\dotsb+a_j\leq N - (j-1) - s - e \leq N - (j-1)$.
Therefore, $|I| = \sum_{1\leq j\leq \lfloor{\frac{N+1}{2}}\rfloor} {\big(\sum_{i=1}^{N - (j-1)}{p_{n,j}}\big)}$. 
For $N = 10$, we have $|I| = 55$. For $N = 64$, we have $|I| = 2,012,557$.
Therefore, $|I|$ is of a size that is computationally reasonable for a range of reasonable block sizes.

\section{Infinite Sequence of Blocks}

\subsection{Derivation of Expected Value of the Length of Consecutive Losses After Erasure Coding}

When multiple blocks are introduced, the ways in which to have consecutive losses becomes more complicated. To see why, let's look at an example:\\

\begin{tiny}
\underline{\hspace{0.15cm}} \underline{\hspace{0.15cm}} \underline{\hspace{0.15cm}} \underline{\hspace{0.15cm}} 
\underline{\hspace{0.07cm}}\underline{X}\underline{\hspace{0.07cm}}	\big|
\underline{\hspace{0.07cm}}\underline{X}\underline{\hspace{0.07cm}} 
\underline{\hspace{0.07cm}}\underline{X}\underline{\hspace{0.07cm}} 
\underline{\hspace{0.07cm}}\underline{X}\underline{\hspace{0.07cm}} 
\underline{\hspace{0.07cm}}\underline{X}\underline{\hspace{0.07cm}}
\underline{\hspace{0.15cm}}							\big|
\underline{\hspace{0.15cm}} \underline{\hspace{0.15cm}} \underline{\hspace{0.15cm}}
\underline{\hspace{0.07cm}}\underline{X}\underline{\hspace{0.07cm}}
\underline{\hspace{0.15cm}} 							\big|
\underline{\hspace{0.15cm}}			
\underline{\hspace{0.07cm}}\underline{X}\underline{\hspace{0.07cm}}
\underline{\hspace{0.07cm}}\underline{X}\underline{\hspace{0.07cm}}
\underline{\hspace{0.07cm}}\underline{X}\underline{\hspace{0.07cm}}	
\underline{\hspace{0.15cm}}							\big|
\underline{\hspace{0.15cm}}  			
\underline{\hspace{0.07cm}}\underline{X}\underline{\hspace{0.07cm}} 
\underline{\hspace{0.07cm}}\underline{X}\underline{\hspace{0.07cm}}
\underline{\hspace{0.15cm}} \underline{\hspace{0.15cm}} 
\\
\end{tiny}

In this example, the bars are used to delineate the boundaries of a block. We have 5 blocks in this example, and there is a consecutive loss of length 5 that spans the first and second blocks. The fact that consecutive losses can span multiple blocks is something we will have to take into account when calculating the expected value.\\

Indexing the notation from the single block case, let $A_i =  (s_i, a_{1_i}, a_{2_i}, ..., a_{j_i}, e_i)$ denote the loss vector for block $i$ where $j_i$ denotes the number of loss rows for block $i$.\\

\textbf{Definition: } The total number of losses for a singl- block loss vector $A_i =  (s_i, a_{1_i}, a_{2_i}, ..., a_{j_i}, e_i)$ is denoted $|A_i|$. That is, $|A_i| = \sum\limits_{k=1}^{j}a_{k_i}$\\

\textbf{Definition: }A $\textit{multiblock pattern of size k}$ is a sequence of loss vectors $(A_1, A_2, \dots , A_k)$ over k blocks, where each block in the sequence has at least one loss each, i.e., $|A_i| \geq 1$ for $i=1, 2, \dots, k$.\\

\textbf{Definition: } Let $M_i$ be the set of all multiblock patterns possible in $i$ number of blocks.\\

\textbf{Definitions: } For a particular mutliblock pattern $(A_1, A_2, \dots, A_M)$, let $T(A_1, \dots,  A_i)$ be the \textit{total number of losses}, and $R(A_1, \dots,  A_i)$ be the the \textit{total number of loss rows}. 

Then the number of losses per row for the multiblock pattern is given by the notation $\frac{T (A_1, ...,  A_i)}{R (A_1, ...,  A_i)}$.\\

The quantity $T(A_1, ...,  A_i)$ is additive. That is, $T(A_1, ...,  A_i) = \sum\limits_{k = 1}^{i} |A_k|$. On the other hand, $R(A_1, ...,  A_i)$ is not the sum of the loss rows for the individual blocks. Rather, $R(A_1, ...,  A_i) = (j_1 + j_2 + \dots + j_i)  - \sum\limits_{i=1}^{i-1}(1 - e_j)(1 - s_{j+1})$. The presence of the summation can be understood by noting that the number of total loss rows will be the number of loss rows in each of the individual blocks minus 1 for each loss row that continues onto another block (because we wish to avoid double counting a loss row everytime it continues onto another block).\\ 

Let $C$ be the average length of loss rows for such an infinite sequence of patterns, and let $E[C]$ be its expectation. \\
\begin{theorem}
Let $E[C]$ denote the expected value for the length of consecutive losses for an infinite number of blocks. Then, \\
\begin{small}
$E[C]= \sum\limits_{i=1}^{\infty} \mkern10mu\mathclap{\sum\limits_{\mkern99mu\text{\((A_1, ... , A_i\))} \in M_i}}\hspace{0.2cm}\frac{T (A_1, ...,  A_i)}{R (A_1, ...,  A_i)}  P(A_1)P(A_2)\dotsm P(A_i)Q(0)$\end{small}\\ 
where $P(A_k) = P(s_k, a_{1_k}, a_{2_k}, \dots, a_{r_k}, e_k)$ is given by Theorem 2 and $Q(0)$ is given by Theorem 1.\\
\end{theorem}

\begin{IEEEproof}
Any pattern of losses and receipts over an infinite sequence of blocks can be viewed as a sequence of multiblock patterns separated by one or more blocks containing all receipts. Since the average length of loss rows for different multiblock patterns in the sequence are independent and identically distributed random variables, the average length of loss rows for a single multiblock pattern will have the same expected value as $C$. We therefore have \\
$E[C]=\sum\limits_{i=1}^{\infty} \mkern15mu\mathclap{\sum\limits_{\mkern80mu\text{\((A_1, ... , A_i\))} \in M_i}}\hspace{0.2cm} E\Big[\frac{T (A_1, ...,  A_i)}{R (A_1, ...,  A_i)} \Big| (A_1, ... ,  A_i), |A_{i+1}|=0\Big]\\ \indent\hspace{3cm}\cdot P((A_1, ... , A_i),|A_{i+1}|=0)\\
\indent\hspace{0.32cm}= \sum\limits_{i=1}^{\infty} \mkern15mu\mathclap{\sum\limits_{\mkern80mu\text{\((A_1, ... , A_i\))} \in M_i}}\hspace{0.2cm}\frac{T (A_1, ...,  A_i)}{R (A_1, ...,  A_i)}\cdot  P((A_1, ... , A_i), |A_{i+1}|=0)$\\
The statement of Theorem 4 follows because the loss vectors of different blocks are independent of one another under the Bernoulli model of network losses.\\
\end{IEEEproof}

One might ask if it is necessary to calculate the series to an infinite number of terms. We will show that if we truncate the sum at some $i=n$, we can get a small bound on the difference between this truncated sum and $E[C]$. We have\\
\begin{small}
$\Bigg|E[C] -  \sum\limits_{i=1}^{n} \mkern10mu\mathclap{\sum\limits_{\mkern99mu\text{\((A_1, ... , A_i\))} \in M_i}}\hspace{0.2cm}\frac{T (A_1, ...,  A_i)}{R (A_1, ...,  A_i)}  P(A_1)P(A_2)\dotsm P(A_i)Q(0)\Bigg|\\
=  \sum\limits_{i=n+1}^{\infty} \mkern10mu\mathclap{\sum\limits_{\mkern99mu\text{\((A_1, ... , A_i\))} \in M_i}}\hspace{0.2cm}\frac{T (A_1, ...,  A_i)}{R (A_1, ...,  A_i)}  P(A_1)P(A_2)\dotsm P(A_i)Q(0)\\
\leq \sum\limits_{i=n+1}^{\infty} \mkern10mu\mathclap{\sum\limits_{\mkern99mu\text{\((A_1, ... , A_i\))} \in M_i}}\hspace{0.2cm}N\cdot i  P(A_1)P(A_2)\dotsm P(A_i)Q(0)\\    
= \sum\limits_{i=n+1}^{\infty}\mkern20mu\mathclap{ N\cdot i Q(0)} \mkern50mu\mathclap{\sum\limits_{\mkern100mu\text{\((A_1, ..., A_i\))} \in M_i}}\hspace{0.15cm} P(A_1)P(A_2)\dotsm P(A_i)\\
= NQ(0)\sum\limits_{i=n+1}^{\infty} i (1-Q(0))^i \\
= NQ(0)\sum\limits_{i=0}^{\infty} i (1-Q(0))^i - NQ(0)\sum\limits_{i=0}^{n} i (1-Q(0))^i \\\\ $\end{small}
$= \frac{N(n+1)(1-Q(0))^{n+1} -Nn(1-Q(0))^{n+2}}{Q(0)}.$

This is typically a small value when $n$ itself is small. Therefore, we can closely approximate $E[C]$ by truncating the upper bound index in the first summation of Theorem 4. We can decrease this bound further, but for brevity we do not include that result here.\\

We include a table below of how large $n$ needs to be to have an error of less than 0.005 in estimating the expected value of the length of consecutive loss with $E[C]$ for various values of $p$, assuming a block size of 10 and redundancy size of 3.

\begin{center}
\begin{tabular}{|c | c|} 
\hline
 $p$ & $n$ \\ [0.5ex] 
 \hline
1\% & 1\\
\hline
5\% & 1\\
\hline
10\% & 2\\
\hline
15\% & 4\\
\hline
25\% & 11\\
\hline
40\% & 64\\
\hline
50\% & 281\\
\hline
60\% & 1947\\
\hline
70\% & 27406\\
\hline
80\% & 1355202\\
\hline
90\% & 1332794850\\
\hline
\end{tabular}
\end{center}
\indent\hspace{1cm}Table 1. Number of terms in series required \\\indent\hspace{2cm}for error less than 0.005\\

Notice that if $p$ is very large, $n$ needs to be very large to get an absolute error of less than $0.005$. However, for the study that motivated this work $[2]$, network loss rates of 15\% and below were most relevant. This is because erasure coding was shown to provide a substantial improvement in voice quality for network loss rates in that range when the block size $N$ and redundancy size $K$ were also in a practical range. For network loss rates in that range, a small value of $n$ is sufficient. Nevertheless, we have derived a recursive $O(N^3)$ algorithm to calculate the expected value of the length of consecutive loss when the $n$ needed to get a close approximation is very high. For brevity we do not include it in this paper.\\

Figure 1 below shows the approximated values of $E[C]$ obtained using our formula with a block size of 5 and redundancy size of 2, both with and without erasure coding. The expected length of consecutive loss under the Bernoulli model without erasure coding is given by $(1-p)^{-1}$.
\includegraphics[scale=0.5]{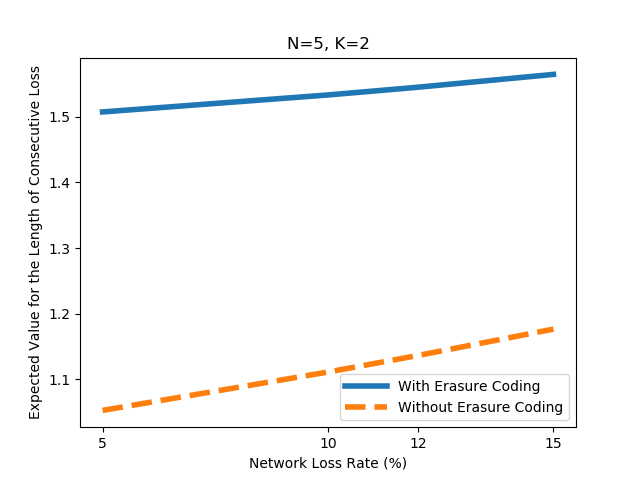}
$$\text{Figure 1. Expected length of consecutive losses comparison}$$

As one would expect, the higher the network loss rate, the greater the expected value of the length of consecutive loss. On the other hand, Figure 1 shows that erasure coding increases the expected length of consecutive loss relative to the expected length without erasure coding. That result initially surprised the authors. However, upon further consideration, this makes sense. We discuss it further in the conclusion.

\section{Conclusion}
Theorem 4 provides an exact expression for estimating the number of losses in a row for a block erasure code as a function of network loss probabilities. Although it requires summing over a sample space that can be very large in principle, we show that it can be applied with low complexity in practice except when network loss rates are very high (or block sizes are very large). At network loss rates for which the model’s computations become infeasible, the number of redundancy packets per block required for any reasonable level of media quality would be so large as to be impractical.\\

Our results provide the new insight that the use of block erasure coding increases the expected number of losses in a row (the measure of loss burstiness required as input by the ITU-T E-Model for voice quality estimation). That result might seem paradoxical since erasure coding only decreases or leaves unchanged the lengths of rows of consecutive losses compared with their lengths if erasure coding were not performed. The paradox is resolved by realizing that the probability erasure coding will recover packets that have been lost in the network from a block decreases as the number of lost packets from the block increases. Hence, the blocks for which there are unrecovered packets after erasure coding will tend to be those with more losses, and therefore with more consecutive losses. The benefits for media quality from the reduction in loss rates from erasure coding will therefore be tempered by the property that unrecovered losses will tend to be the losses most concentrated together, and hence the most detrimental to intelligibility. Nevertheless, the study in [2] shows that the net benefit of erasure coding for voice quality will generally be large, even accounting for added delays.

Although the model developed here assumes Bernoulli losses in the network, much of its machinery would apply for any model of network loss. Hence, our analysis can serve as a foundation for future extensions to the current model.

\bibliographystyle{IEEEtran}
\bibliography{IEEEabrv,mybibfile}
\section{References}
\noindent $[1]$  Rizzo, Luigi. (1999). Effective Erasure Codes for Reliable Computer Communication Protocols. ACM SIGCOMM Computer Communication Review. 27. 10.1145/263876.263881.\\
$[2]$  A. David, A. DeSimone, K. Fendick, S. Handy, B. McCann. Improving Conversation Quality for VoIP Through Block Erasure Coding, submitted to IEEE ICC 2020\\
$[3]$ Chan, K.-S., Yeung, L. K., and Shao, W. (2005). Contention-based MAC protocols with erasure coding for wireless data networks. Ad Hoc Networks, 3(4), 495–506. doi: 10.1016/j.adhoc.2004.02.003\\
$[4]$ Cook, John and Primmer, Robert and Kwant, Ab. (2013). Compare cost and performance of replication and erasure coding. Hitachi Review. 63. \\

\end{document}